\documentclass[aps,prl,twocolumn,nofootinbib,floatfix]{revtex4-2}

\usepackage{amsmath,amssymb,bm}
\usepackage{graphicx}
\usepackage{epstopdf}
\usepackage{amsfonts}
\usepackage{amssymb}
\usepackage{amsbsy}
\usepackage{amsmath}
\usepackage{latexsym}
\usepackage{sansmath}
\usepackage{subfigure}
\usepackage{lipsum}
\usepackage{float}
\usepackage{bm}
\usepackage{color}
\usepackage{comment}
\usepackage{csquotes}
\usepackage{physics}
\usepackage{eufrak}
\usepackage{accents}
\usepackage[colorlinks=true,breaklinks]{hyperref}
\usepackage[plain]{algorithm}
\usepackage{algpseudocode}

\newcommand{\pd}{\partial}
\def\mH{\mathcal{H}}
\def\ga{\gamma}
\def\be{\begin{equation}}
\def\ee{\end{equation}}
\def\bea {\begin{eqnarray}}
\def\eea {\end{eqnarray}}
\def\nn {\nonumber}

\def\dd{{\rm d}}
\def\mE{\mathcal{E}}
\def\f{\frac}
\def\ea{{E}^{a}}
\def\eb{{E}^{b}}

\def\lp{\ell_{\rm Pl}}

\begin{document}
 
\title{Quantum gravity of dust collapse: shock waves from black holes}

\author{Viqar Husain} \email{vhusain@unb.ca}
\affiliation{Department of Mathematics and Statistics, University of New Brunswick,
Fredericton, NB, Canada E3B 5A3}

\author{Jarod George Kelly} \email{jarod.kelly@unb.ca}
\affiliation{Department of Mathematics and Statistics, University of New Brunswick,
Fredericton, NB, Canada E3B 5A3}

\author{Robert Santacruz} \email{robert.santacruz@unb.ca}
\affiliation{Department of Mathematics and Statistics, University of New Brunswick,
Fredericton, NB, Canada E3B 5A3}

\author{Edward Wilson-Ewing} \email{edward.wilson-ewing@unb.ca}
\affiliation{Department of Mathematics and Statistics, University of New Brunswick,
Fredericton, NB, Canada E3B 5A3}

\begin{abstract}
 
We study the quantum gravitational collapse of spherically symmetric pressureless dust. Using an effective equation derived from a polymer quantization in the connection-triad phase space variables of general relativity, we find numerically, for a variety of initial dust configurations, that (i) trapped surfaces form and disappear as an initially collapsing density profile evolves into an outgoing shockwave; (ii) black hole lifetime is proportional to the square of its mass; and (iii) there is no mass inflation at inner apparent horizons. These results provide a substantially different view of black hole formation and subsequent evolution than found from semiclassical analyses.

\end{abstract}

\maketitle

In classical general relativity, black holes are the final state of the gravitational collapse of sufficiently massive configurations of matter. Hints of the stability of black holes first appeared in linear stability analyses of black hole spacetimes, and subsequently through analytical \cite{Christodoulou:1987vu} and numerical studies \cite{Choptuik:1992jv} of scalar field collapse in spherical symmetry. 

In the semiclassical gravity context of quantum test fields on curved spacetime, it was discovered that black holes radiate and lose mass \cite{Hawking:1975vcx}. The effect of the radiation on black holes is usually modelled by a sequence of quasi-static configurations of shrinking mass $M(t)$. Since Hawking radiation is that of a black body, at least to a first approximation, mass loss is described by the Stefan-Boltzmann law together with the facts that black hole mass is proportional to its radius, and its temperature is inversely proportional to mass. This implies a black hole lifetime proportional to the cube of its initial mass $\sim M^3/m_{\rm Pl}^2$. Thus while black holes are stable in classical gravity, they are not in the semiclassical regime. 

This lifetime result assumes that semiclassical gravity holds up to the final stages of black hole evaporation. However, since a black hole gets hotter and the curvature at the horizon gets larger as its mass shrinks, the semiclassical approximation is expected to fail at least in the late stages of evaporation due to the ``backreaction" of Hawking radiation on the spacetime, beyond just the effect due to shrinking mass. 

What is required for a complete understanding of black hole physics is nothing less than a unified description of gravitational collapse in quantum gravity that can describe the entire evolution, from the initial collapse to the late stages of Hawking radiation and subsequent evolution of matter and spacetime. Such an understanding might arise from the quantization of a classical model like pressureless dust or scalar field in spherical symmetry. 

A variety of studies use spherically symmetric models with the ultimate aim of understanding the collapse process and subsequent evolution quantum mechanically. There are two main approaches, those that start with a black hole spacetime, quantize its interior as a cosmological spacetime, and match it to an exterior Schwarzschild metric \cite{Modesto:2008im, Bohmer:2007wi, Gambini:2013ooa, Corichi:2015xia, Ashtekar:2018cay, Bodendorfer:2019cyv, BenAchour:2020bdt}, and those that avoid such an interior-exterior separation (and do not assume a pre-exisiting horizon) \cite{Husain:2004yy, Hayward:2005gi, Campiglia:2007pr, Hossenfelder:2009fc, Kreienbuehl:2010vc, Ziprick:2009nd, Bojowald:2018xxu, Benitez:2020szx, BenAchour:2020gon, Gambini:2020nsf, Kelly:2020uwj, Munch:2020czs}. This work follows the second approach.

We develop and study an effective quantum gravity formalism for pressureless dust collapse in spherical symmetry. Classically this is the Lema\^itre-Tolman-Bondi (LTB) spacetime. Earlier work towards a quantum theory of this model appears in \cite{Vaz:2011zz, Bojowald:2008ja, Kiefer:2019csi, Kelly:2020lec, Han:2020uhb, Giesel:2021dug}. We derive effective quantum-gravity corrected equations in connection-triad variables \cite{Ashtekar:2004eh, Bodendorfer:2016uat}, and solve them numerically for two classes of initial data. Our main result is that in-falling dust bounces when spacetime curvature reaches the Planck scale, and forms an outgoing shock wave. When the shock wave reaches the Schwarzschild radius, the trapped region(s) formed during collapse disappear. This rules out remnants, no-horizon formation results such as in \cite{Kawai:2013mda, Baccetti:2017ioi}, and realizes explicitly heuristic ideas concerning the formation and evolution of non-singular black holes discussed in the literature \cite{Hayward:2005gi, Ashtekar:2005cj, Rovelli:2014cta}. We calculate the lifetime of a black hole to be $T \sim M^2$, a time much shorter than the $T\sim M^3$ predicted by the semiclassical approximation.

The class of metrics we study is 
\be \label{LTB}
\dd s^2 = - \dd t^2 + \f{\big( \dd x + N^x(x,t) \dd t \big)^2}{1 + \mE(x,t)} + x^2 \dd \Omega^2,
\ee
where $t\in \mathbb{R}, \ x\in \mathbb{R}_{\ge0}$ and $\dd \Omega^2$ is the unit sphere metric. This is a Painlev\'e-Gullstrand form of the spherically symmetric LTB spacetimes. The stress-energy tensor for dust is $T_{\mu\nu} = \rho(x,t) \, u_\mu u_\nu$ where $u^\mu$ is the unit 4-velocity of the dust field and $\rho(x,t)$ its mass density.

Motivated by loop quantum gravity, we use the triad-connection phase space for general relativity that arises from the Holst action \cite{Holst:1995pc}, minimally coupled to dust:
\begin{align}
S =& \, \f{1}{16 \pi G} \! \int \dd^4x \, |e| e^\mu_I e^\nu_J \left( F_{\mu\nu}{}^{IJ} + \f{1}{2\ga} \epsilon^{IJ}{}_{KL} F_{\mu\nu}{}^{KL} \right) \nn \\
& -\f{1}{2} \int \dd^4x \, |e| \, \rho \left( e^\mu_I e^\nu_J \eta^{IJ} \partial_\mu T \partial_\nu T + 1 \right),
\end{align}
where $e^\mu_I$ are the tetrads, $e$ is their determinant, $F_{\mu\nu}{}^{IJ}$ is the field strength of the $SO(3,1)$ connection $w_\mu{}^{IJ}$, $T$ is the dust field with mass density $\rho$, and $\eta^{IJ}$ is the (inverse) Minkowski metric. The Hamiltonian decomposition of this action gives the canonically conjugate gravitational variables $A_a^i = \epsilon^i{}_{jk} w_a{}^{jk} + \ga w_a{}^{0i}$ (the Ashtekar-Barbero connection) and $E^a_i = |e| e^a_i$ (the densitized triads) \cite{Holst:1995pc}. (Here $\mu,\nu,\cdots$ are spacetime indices, $a,b,\cdots$ are spatial indices, $I,J,\cdots$ are Lorentz indices, and $i,j,\cdots$ are their spatial components.)

For spherically symmetric spacetimes \cite{Bojowald:2005cb, Kelly:2020uwj} (with $a$ and $b$ now as labels),
\begin{align}
E^x_1 &= E^a \sin\theta, & E^\theta_2 &= E^b \sin\theta, & E^\phi_3 &= E^b, \nn \\
A_x^1 &= a, & A_\theta^2 &= b, & A_\phi^3 &= b \sin\theta, \\
A_\theta^3 &= - \f{\partial_x E^a}{2 E^b}, & A_\phi^2 &= \f{\partial_x E^a}{2 E^b} \sin\theta, & A_\phi^2 &= -\cos\theta; \nn
\end{align}
all other components vanish. The basic Possion bracket relations are
\bea
\{a(x,t), E^a(x',t)\} &=& 2 \, G \ga ~ \delta(x - x'), \nn\\
\{b(x,t), E^b(x',t)\} &=& G \ga ~ \delta(x - x'), \\
\{T(x,t),p_T(x',t) \} &=& \delta(x - x')/4\pi, \nn
\eea
where $p_T$ is the momentum of $T$, see \cite{Husain:2011tk} for details.

These variables are subject to the scalar and diffeomorphism constraints of general relativity. We fix the gauge freedom generated by the first constraint by setting $T=t$ \cite{Husain:2011tk}, and the second by setting $\ea = x^2$; this is the standard areal gauge \cite{Kuchar:1994zk} that gives $1/(1+\mE) = (\eb/x)^2$ in the metric \eqref{LTB}. Preservation of these gauge-fixing conditions under evolution requires $N=1$ \cite{Husain:2011tk, Kelly:2020lec} and $N^x = -b/\gamma$ \cite{Kelly:2020uwj}. So gauge-fixed, the Hamiltonian theory has phase space variables $(b, E^b)$, and the reduced canonical action (dropping boundary terms) is
\begin{align}
\label{eq:redH}
S_{GF} &= \int \dd t \int \dd x \, \left( \f{\dot{b}\eb}{G\gamma} - \mathcal{H}_{\rm phys} \right), \\
\label{Hphys}
\mH_{\rm phys} &\equiv - \, \frac{1}{2G\gamma} \biggr[ \frac{\eb}{\gamma x} \pd_x( x b^2 ) + \frac{\gamma \eb}{x} 
+ \frac{\gamma x}{\eb} \biggr].
\end{align}
$\mH_{\rm phys}$ is the true physical Hamiltonian. The dust energy density is 
\be \label{density}
\rho = - ~ \f{\mathcal{H}_{\rm phys}}{4 \pi x \, \eb},
\ee
and the total mass contained within a radius $x$ is 
\be
m(x,t) = 4\pi \int_0^x \! \dd \tilde x \, \tilde x \, E^b \rho = - \int_0^x \dd \tilde x \, \mH_{\rm phys}.
\ee 
Exactly the same $\mH_{\rm phys}$ can be derived from the Einstein-Hilbert action coupled to dust, assuming the metric has the gauge-fixed form given above.

We quantize this theory by first defining a discretization of the classical theory on a radial lattice, a procedure similar in spirit to lattice gauge theory. For simplicity we consider an equispaced radial lattice $x \rightarrow x_n, n=0\cdots N$; $x_{n+1}-x_n = w$; $f_n = f(x_n)$; and $\pd_x f(x_n) \rightarrow (f_{n+1}-f_n)/w$; (a non-equispaced radial lattice could also be used). The phase space variables are then defined at discrete points $b_n(t) \equiv b(x_n,t)$ and $E^b_n(t) \equiv E^b(x_n,t)$, and their fundamental Poisson bracket is $\{b_n, E^b_m \} = G\gamma\delta_{m,n}/w$. The physical Hamiltonian $\mathcal{H}_{\rm phys}$ becomes a sum over lattice sites.

In general a quantum theory depends on the choice of fundamental variables and Hilbert space. Our choice is motivated by loop quantum gravity; the fundamental operators are geometrical, and polymer quantization ensures background independence as the Hilbert space has an inner product independent of the metric (itself subject to quantum fluctuations). The polymer quantization is defined by representing at each point $x_n$ the algebra
\be \label{algebra}
\{\exp(i\mu b_n ), E^b_n \} = \f{i\mu G\gamma}{w} \exp(i\mu b_n )
\ee
on the Hilbert space $\mathbb{H}_n$ with basis vectors $|E^b\rangle_n$ and inner product $ \ _n\langle E^b| E^{b\prime} \rangle_n = \delta _{E^b ,E^{b\prime}}$; this is similar to the momentum representation for a particle on $\mathbb{R}$ with the difference that the r.h.s.\ is the Kronecker delta. The discrete inner product is a key feature of the polymer Hilbert space. With ${\cal N}_n(\mu) \equiv \exp(i\mu b_n)$, the representation is
\bea
\hat{E}^b_n | E^b\rangle_n &=& E^b | E^b\rangle_n, \nn\\
\hat{\cal N}_n(\mu) | E^b\rangle_n &=& | E^b + \hbar G\gamma \mu/w\rangle.
\label{elemOP}
\eea
The Hilbert space for the entire lattice is the tensor product $\displaystyle \mathbb{H} = \otimes_{n=0}^N\ \mathbb{H}_n$.

The operator for the Hamiltonian $\mathcal{H}_{\rm phys}$ requires a definition of $\hat{b}_n$ from the elementary operators \eqref{elemOP}. The simplest self-adjoint possibility is (see, e.g., \cite{Bodendorfer:2016uat})
\be \label{b-op}
\hat{b}_n(\mu) = \frac{1}{2i\mu} \left( \hat{\cal N}_n(\mu) - \hat{\cal N}_n(\mu)^\dagger \right).
\ee
While other choices are possible here (e.g., by adding higher powers of $\hat{\cal N}_n$), the key feature of any choice is that $\hat{b}_n(\mu)$ is a bounded operator. The parameter $\mu$ is as yet unspecified---it may be a constant, or a function of $x_n$ and/or $E^b_n$ without affecting the algebra; fixing it requires a physical input. To do this, recall that $E^b$ is the $\theta$ component of the triad, and ${\cal N}_n(\mu)$ generates translations in the $\theta$ direction by an angle $\mu$. The physical translation distance at a radius $x_n$ is $|\Delta s| = x_n \mu = \ell_{\rm Pl}$, where the last equality is the required physical input: the elementary translation operator in the Hamiltonian should correspond to a Planck-length step. This sets \cite{Kelly:2020uwj}
\be \label{mun}
\mu = \frac{\ell_{\rm Pl}}{x_n}.
\ee 
This summarizes the quantization; details appear in \cite{wip}.

An effective Hamiltonian can be extracted from the quantization by replacing $b_n$ in the discretized $\mH_{\rm phys}$ by the corresponding classical function that contains Planck length corrections, and then taking the continuum limit. Using \eqref{b-op} and \eqref{mun} this classical function is 
\be \label{bn-eff}
\tilde{b}_n = \frac{x_n}{\ell_{\rm Pl}} \sin\left( \frac{\ell_{\rm Pl} b_n}{x_n} \right), 
\ee
and the continuum limit gives
\be \label{Heff}
\mathcal{H}_{\rm phys}^{\rm eff} \! =- \f{1}{2G\gamma} \! \left[ \f{\eb}{\gamma \ell_{\rm Pl}^2 x} \pd_x \! \left( \! x^3 \sin^2 \f{\ell_{\rm Pl} b}{x} \right)
+ \f{\gamma x}{\eb} + \frac{\gamma \eb}{x} \right].
\ee
The key difference between the classical Hamiltonian \eqref{Hphys} and this effective one is that here the extrinsic curvature $b$ is contained in a bounded function. This is the feature responsible for resolving the singularity. The effective equations derived from (\ref{Heff}) capture spacetime discreteness at the Planck scale through this bounded function; it is an approximation where quantum fluctuations are small and spacetime geometry is well approximated by a continuous metric \cite{Rovelli:2013zaa}.

For numerical calculations we set $\gamma=\ell_{\rm Pl}=1$, and the effective equations obtained from this Hamiltonian are 
\be \label{beff}
\dot b = \f{x}{2 E_b^2} - \f{1}{2x} - x \sin \f{b}{x} \left[ \f{3}{2} \sin \f{b}{x} + x \, \partial_x \sin \f{b}{x} \right],
\ee
\be \label{Ebeff}
\dot E_b = - \, \frac{x^2}{2} \, \partial_x \left(\f{E_b}{x}\right) \sin \f{b}{x} ~ \cos \f{b}{x}.
\ee
Clearly $E^b=x$ is a solution of the second equation; in classical general relativity these are the LTB spacetimes with ${\cal E}(x) =0$ in the metric \eqref{LTB}. In the remainder of this Letter, we consider this class of spacetimes and leave the study of solutions with $E^b \neq x$ for future work. With $E^b=x$, the effective equation \eqref{beff} simplifies,
\be \label{eomb}
\dot b + \f{1}{2 x} \pd_x \left( x^3 \sin^2 \f{b}{x} \right) = 0.
\ee

To summarize, after gauge-fixing the scalar and diffeomorphism constraints, we quantized the LTB theory; from this we obtained the effective equations \eqref{beff}-\eqref{Ebeff} with $\hbar$ corrections, and then restricted to the class of solutions with $E^b=x$, resulting in \eqref{eomb}.

The effective equations \eqref{beff}-\eqref{Ebeff} reduce to the classical ones when $b \ll x/\lp$. Quantum gravity corrections are large only where the extrinsic curvature $b$ nears the Planck scale---this occurs at the bounce point and at the shock wave discontinuity, regardless of how far outward the shock wave has propagated. This is how quantum gravity effects appear at macroscopic radial distances.

For the class of solutions with $E^b=x$, the effective metric is $\dd s^2 = -\dd t^2 + (\dd x + N^x \dd t)^2 + x^2 \dd\Omega^2$, and
\be \label{q-density}
\rho = -\f{\mH_{\rm phys}^{\rm eff}}{4 \pi x^2}
= \f{1}{8 \pi x^2} \, \pd_x \left( x^3 \sin^2 \f{b}{x} \right).
\ee

Recalling that $N^x=-b$ is induced by the gauge-fixing condition $E^a=x^2$, we require an effective expression for $N^x$ compatible with the one for $b^2$ in $\mH_{\rm phys}$ \eqref{Hphys}. This is 
\be
\label{Nxeff}
N^x = -\f{x}{2}\sin\left(\f{2b}{x}\right),
\ee 
a form also used in vacuum spherical symmetry \cite{Gambini:2020nsf, Kelly:2020uwj}.

\begin{figure*}
\begin{center}
\includegraphics[width=0.82\textwidth]{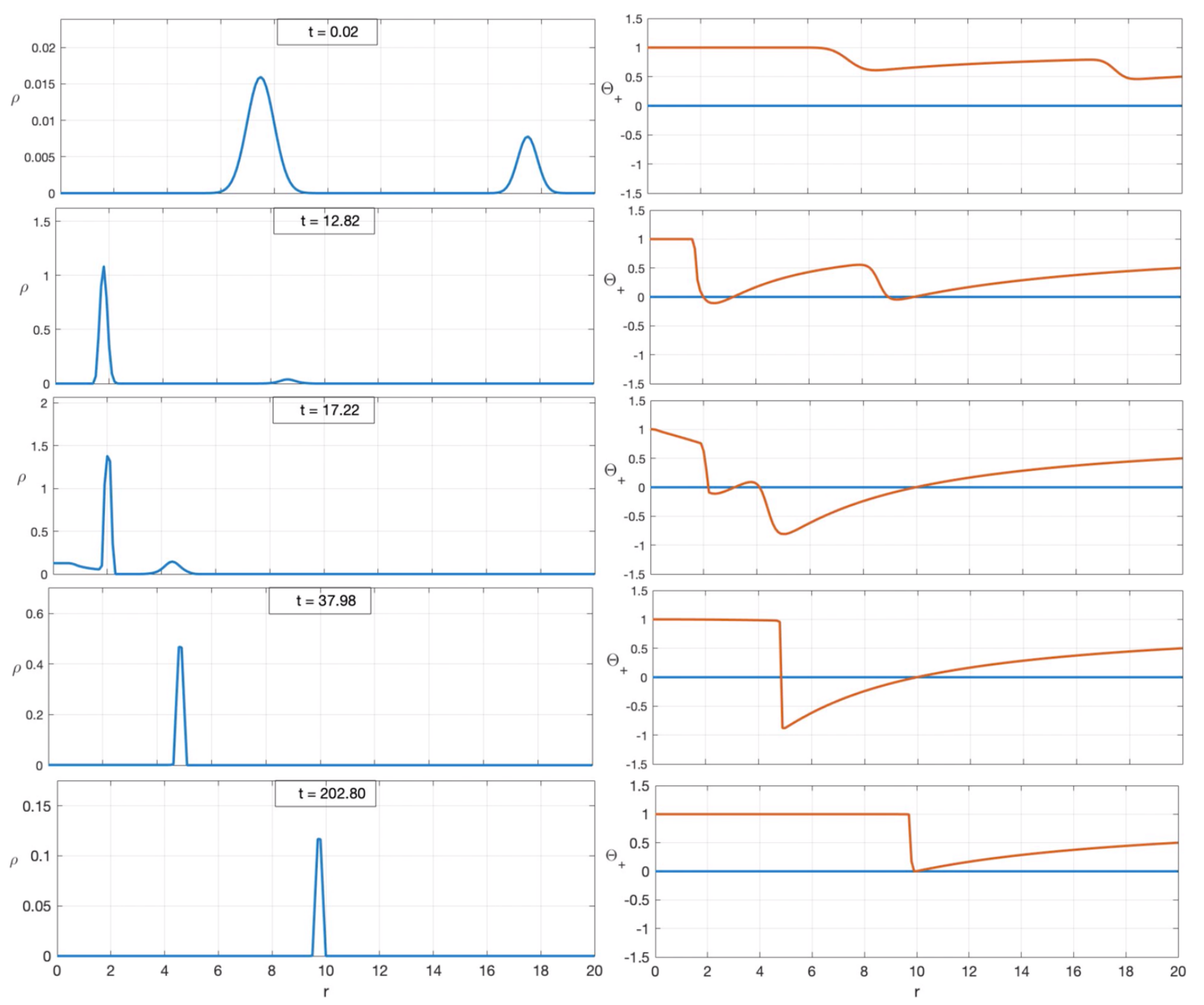}
\caption{Dust density $\rho$ (left column, the vertical scale varies from frame to frame) and apparent horizon function $\Theta_+$ (right column) at the times indicated. The initial data is $\rho_0(r)= \exp[-2(x-1.5M)^2]/5 + \exp[-4(x-3.5M)^2]/10$ with $M=5$ in \eqref{m0}. The first two rows show the collapse, the middle one shows the bounce, and the last two show the outgoing shock wave. Apparent horizons are located at the zeros of $\Theta_+$, in this case the outermost apparent horizon forms and stays at $r=x=2M =10$ until the shock wave reaches it at $t\sim202$.}
\label{2peak}
\end{center}
\end{figure*}

 
To solve \eqref{eomb} numerically, we define $B(x,t) = x \, b(x,t)$ so \eqref{eomb} becomes a conservation law, and then express it as an integral equation to allow discontinuities in $B$,
\be \label{eomB}
\f{\dd}{\dd t} \int_{x_1}^{x_2} \!\! B(x,t) \, \dd x + \frac{1}{2} \left[ x^3 \sin^2 \left(\frac{B(x,t)}{x^2}\right) \right]^{x_2}_{x_1} =0 \, ;
\ee
the term in brackets is the mass function
\be \label{mass-B}
m(x,t) = 4\pi \int_0^x r^2 \rho(r,t)\,\dd r = \frac{1}{2} x^3 \sin^2\left(\frac{B}{x^2}\right).
\ee

Our numerical procedure is the well-known Godunov method: the integration domain is divided into cells, and the Riemann problem is solved in each cell while enforcing flux continuity between cells \cite{Leveque_2002}. The only necessary generalization is due to the current in \eqref{eomB}, which depends on $x$; we address this by evaluating $x$ at the boundary between two cells when computing the flux.

To specify initial data we use two density profiles,
\begin{eqnarray}
\rho_0^G(x) &=& \exp [-(x-x_0)^2/\sigma^2], \label{g-profile} \\
\rho_0^T(x) &=& 1+ \tanh [-(x-x_0)/\sigma]. \label{t-profile}
\end{eqnarray}
The first is a smooth dust ring centred at $x_0$, and the second is a star-like distribution of near constant density and initial radius $x_0$. We rescale these profiles to give a total mass $M$ through
\be
m_0(x) = M \frac{\int_0^x \dd r \, r^2 \rho_0(r)}{\int_0^\infty \dd r \, r^2 \rho_0(r)};
\label{m0}
\ee 
with \eqref{mass-B}, this determines the initial $B_0(x)$. We set $\sigma$ and $x_0$ such that initially there is no apparent horizon.
 
At each time step we compute the density $\rho(x,t)$ and
\be
\Theta_+ = |\grad x|^2 =1-(N^x)^2 = 1- \frac{x^2}{4}\sin^2 \frac{2B}{x^2},
\ee
using the expression \eqref{Nxeff} for the effective $N^x$; the zeros of $\Theta_+$ give the locations of apparent horizons \cite{Faraoni:2016xgy}.

In Fig.~\ref{2peak} are frames from the evolution of a linear combination of initial Gaussian data. We find: (i) during collapse, apparent horizons form in pairs; (ii) there is a bounce near the origin; (iii) an outgoing gravitational shock wave forms, and horizons disappear as the shock moves outward; (iv) the density always remains bounded and the total mass is conserved; (v) there is no instability or mass inflation (unlike in classical general relativity \cite{Eardley:1974zz, Poisson:1989zz}); (vi) after the inner profile bounces, its collision with the second profile does not result in recollapse; (vii) there is no curvature singularity. These features appear in all our simulations and are not affected by $\sigma$ or $x_0$.

\begin{figure}[t]
\begin{center}
\includegraphics[width=0.85\columnwidth]{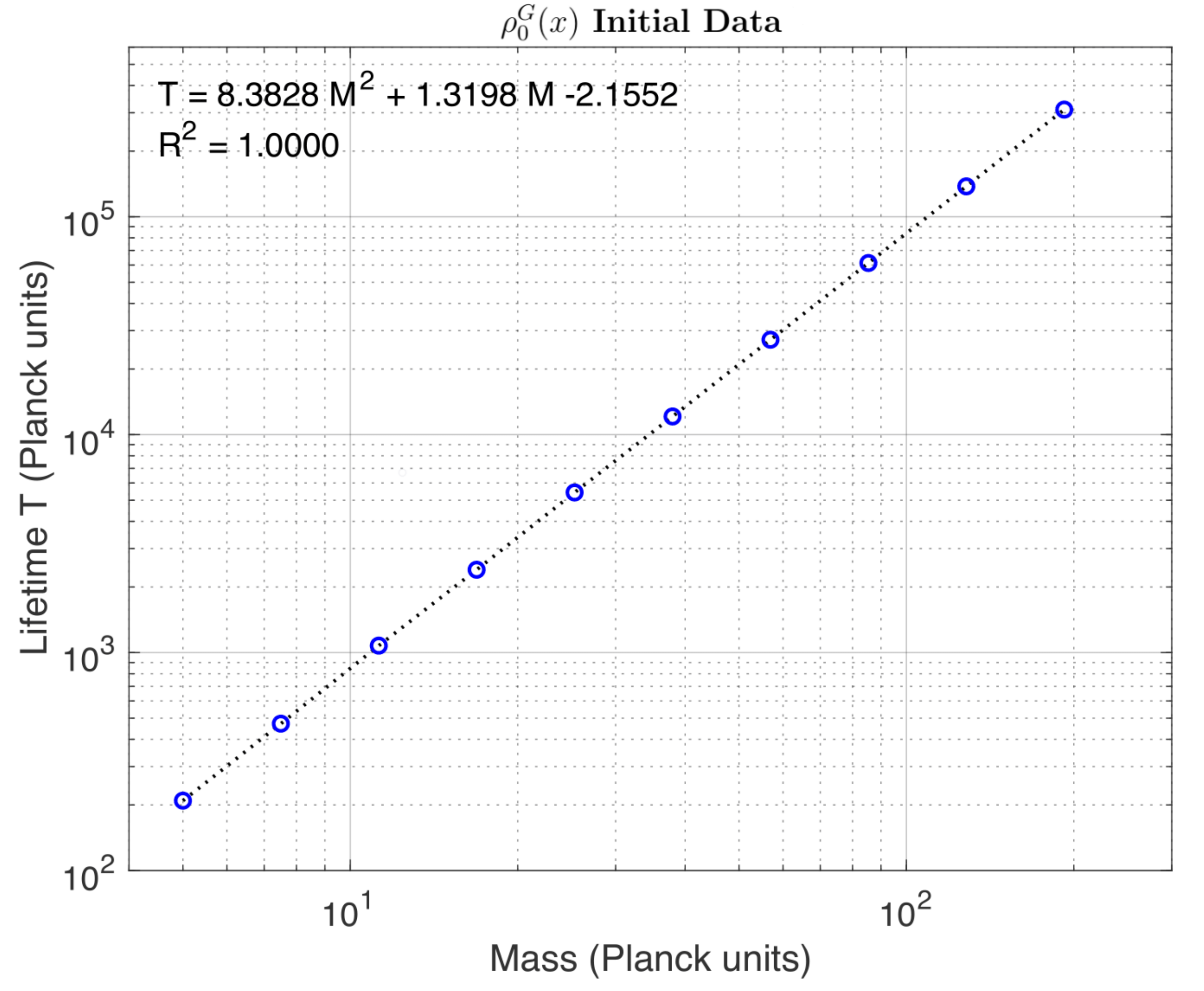}
\caption{Black hole lifetime for initial data \eqref{g-profile} with $\sigma = 0.5$.}
\label{mgauss}
\end{center}
\end{figure}

\begin{figure}[t]
\begin{center}
\includegraphics[width=0.85\columnwidth]{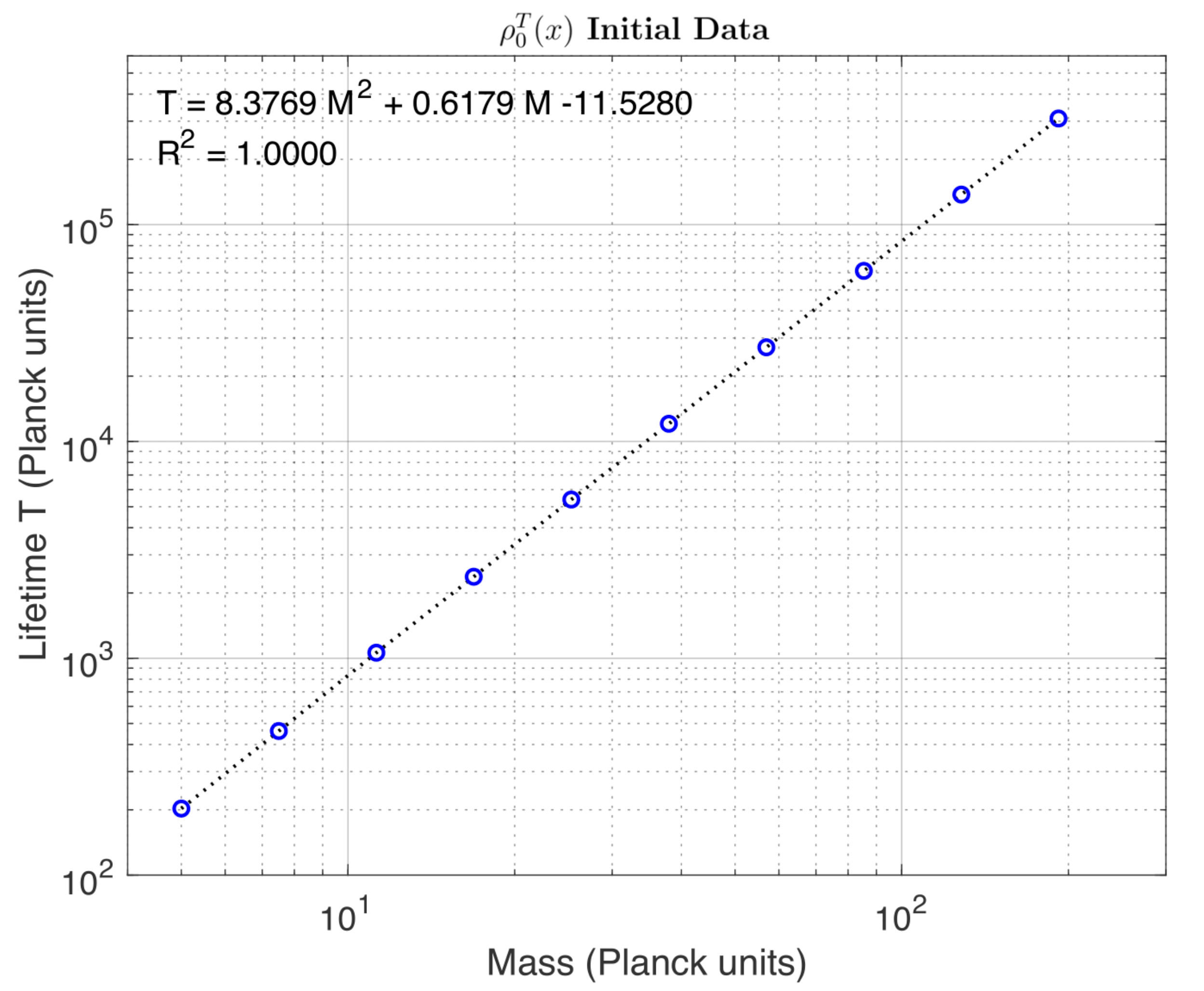}
\caption{Black hole lifetime for initial data \eqref{t-profile} with $\sigma = 0.5$.}
\label{mtanh}
\end{center}
\end{figure}

\begin{figure}[t]
\begin{center}
\includegraphics[width=0.75\columnwidth]{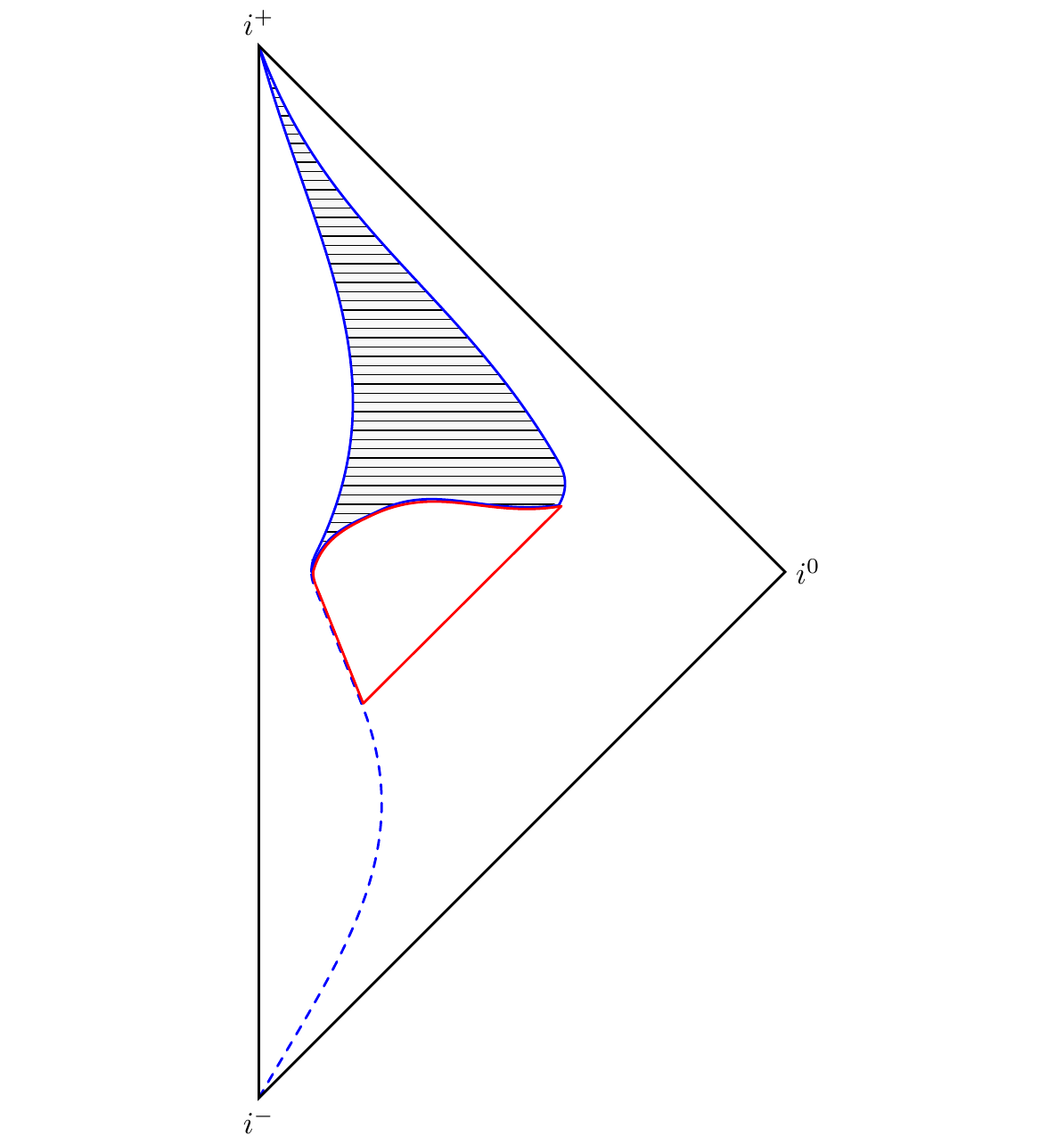}
\caption{Conformal diagram for dust collapse and bounce. Red lines show the outer and inner apparent horizons---the outer horizon becomes and remains null once all in-falling matter has passed through it; the dashed blue line shows a typical ingoing dust trajectory; the solid blue lines show the outgoing shock wave trajectory according to the interior and exterior metrics respectively---these metrics differ due to the shock discontinuity; the shaded portion of the diagram is excised and the two solid blue lines are identified. To an outside observer the outer horizon disappears as the shock wave emerges.}
\label{conformal}
\end{center}
\end{figure}

We computed the black hole lifetime $T$ as a function of the total mass $M$ for the initial density configurations \eqref{g-profile} and \eqref{t-profile} by recording the time between the formation of the outermost apparent horizon and its disappearance, see Figs.~\ref{mgauss} and \ref{mtanh} (for $\sigma=1/2$). The log-log plots show a linear dependence, and the fit (shown in the figures) gives the coefficient of $M^2$ as approximately $8\pi/3$; to leading order $T \sim (8\pi/3)M^2$ over three orders of magnitude in $T$ for both initial data profiles. Finally, Fig.~\ref{conformal} shows the conformal diagram deduced from our simulations.


To summarize, we derived effective Hamiltonian equations that describe quantum gravitational features of dust collapse. Our approach is based on: (i) a complete gauge-fixing of the Hamiltonian and diffeomorphism constraints that gives a physical Hamiltonian, (ii) a discretization and quantization of this system in a polymer framework incorporating a minimal length, and (iii) numerical integration of a subclass of the effective equations. We find numerically that black holes are transitory and non-singular, with a lifetime proportional to $M^2$.

The black hole lifetime result has consequences for Hawking radiation, which will start when the outer apparent horizon forms, and end when it disappears a time $\sim M^2$ later. For $M \gg m_{\rm Pl}$, this is less than the Page time \cite{Page:1993df} when Hawking radiation is maximally entangled with the black hole. This, combined with the absence of an event horizon or singularity, suggests a resolution of the information loss problem: information is recovered after the apparent horizons vanish. We leave an investigation of these questions for future work.

\smallskip

\acknowledgments
\noindent {\it Acknowledgements:} This work was supported by the Natural Sciences and Engineering Research Council of Canada. E.W.-E.~was also supported by the UNB Fritz Grein Research Award.

\raggedright

\end{document}